\documentclass[prl,twocolumn,showpacs,floatfix,amsfonts]{revtex4}
\usepackage{graphicx,graphics,color,epsfig}
\usepackage{bm}
\usepackage{amsmath}
\usepackage{amssymb}
\usepackage{fancyheadings}
\begin{document}
\title{Equation-of-Motion Approach to Dynamical Mean Field
Theory}
\author{Jian-Xin Zhu, R. C. Albers, and J. M. Wills}
\affiliation{Theoretical Division, Los Alamos National Laboratory,
Los Alamos, New Mexico 87545}
\date{\today}
\begin{abstract}
We propose using an equation-of-motion approach as an impurity solver
for dynamical mean field theory. As an illustration of this
technique, we consider a finite-$U$ Hubbard model defined on the
Bethe lattice with infinite connectivity at arbitrary filling.
Depending on the filling, the spectra that is obtained exhibits a
quasiparticle peak, and lower and upper Hubbard bands as typical
features of strongly correlated materials. The results are also
compared and in good agreement with exact diagonalization. We also
find a different picture of the spectral weight transfer than the
iterative perturbation theory.
\end{abstract}
\pacs{71.30.+h, 71.10.Fd, 71.27.+a, 75.10.Lp}
\maketitle

Theoretical understanding of strongly correlated electron systems
has been a central and challenging problem in condensed matter
physics. Dynamical mean field theory
(DMFT)~\cite{AGeorges96,QSi96,SYSavrasov01} has proven to be a
very powerful theoretical tool that is not only tractable but also
flexible enough to include material-specific details into the
calculations. The essence of the DMFT is to map the lattice
problem onto a single impurity problem plus a set of
self-consistency conditions through which the characteristics of a
specific lattice model enters. In particular, all models with only
on-site Hubbard interaction can be mapped onto the single impurity
Anderson model (SIAM)~\cite{PWAnderson61}. The problem is then to
find a suitable solver for the SIAM. For the last few years,
several numerical and analytical techniques to the SIAM have been
explored. They include quantum Monte Carlo
(QMC)~\cite{MJarrell92,MRozenberg92,AGeorges92b}, iterative
perturbation theory
(IPT)~\cite{AGeorges92a,XYZhang93,HKajueter96,MPotthoff97}, and
exact diagonalization (ED)~\cite{MCaffarel94,QSi94}. Each method
has its limitations. Except for some special models~\cite{Zhu03},
QMC has difficulty in accessing very low temperatures, where
statistical and finite time-step discretization errors become
significant. Exact diagonalization, although exact, as the name
implies, is limited by available computer time to a small number
of orbitals, and hence it is difficult to obtain a smooth density
of states. Iterative perturbation theory overcomes the drawbacks
of these other two methods. However, in its original form, IPT  is
only valid at half filling. Away from half filling, it involves an
ansatz and has instability problems.

The aim of this paper is to present an equation-of-motion (EOM)
approach that can overcome the difficulties of QMC and ED while
still maintaining the numerical efficiency and other advantages of
IPT.

For concreteness, we demonstrate the method on the single-band
Hubbard model for a Bethe lattice of infinite
connectivity,
$z\rightarrow \infty$. The Hamiltonian is
\begin{equation}
H=-\frac{t}{\sqrt{z}}\sum_{ij,\sigma} c_{i\sigma}^{\dagger}
c_{j\sigma} +U\sum_{i} n_{i\uparrow} n_{i\downarrow}\;,
\label{EQ:HAMIL_LATT}
\end{equation}
where $t$ (fixed) is the hopping integral for electrons between $z$
neighbors, $n_i$ the occupation number on site $i$,
and $U$ is the
on-site Hubbard interaction.

We study this model using the DMFT approach, which maps the model
onto a self-consistent SIAM as given by the Hamiltonian:
\begin{eqnarray}
H_{\mbox{\small SIAM}}&=& \sum_{k\sigma} \epsilon_{k\sigma}
c_{k\sigma}^{\dagger} c_{k\sigma} + \sum_{k,\sigma} (V_{k\sigma}
c_{k\sigma}^{\dagger} d_{\sigma} +
V_{k\sigma}^{*}d_{\sigma}^{\dagger} c_{k\sigma})\nonumber \\ && +
\sum_{\sigma} \epsilon_{d\sigma} d_{\sigma}^{\dagger} d_{\sigma}
+ Un_{d\uparrow}
n_{d\downarrow}\;,
\label{EQ:HAMIL_SIAM}
\end{eqnarray}
where $c_{k\sigma}^{\dagger}$ ($c_{k\sigma}$) are the creation
(annihilation) operators for the conduction spin-$\frac{1}{2}$
fermionic bath with the energy dispersion $\epsilon_{k\sigma}$ and
$d_{\sigma}^{\dagger}$ ($d_{\sigma}$) for the impurity orbital
with energy $\epsilon_{d\sigma}$, $U$ is the on-site Coulomb
interaction between the electrons on the impurity, and
$V_{k\sigma}$ is the coupling between the bath and impurity
orbital. The effective parameters $\epsilon_{k}$ and $V_{k\sigma}$
enter the hybridization function \begin{equation}
\Delta_{\sigma}(i\omega_{n}) = \sum_{k} \frac{\vert V_{k\sigma}
\vert^{2} }{i\omega_{n} -\epsilon_{k\sigma}}\;, \end{equation}
which is found by solving a self-consistent condition. Here
$\omega_{n}=(2n+1)\pi T$ is the Matsubara frequency for fermions
with $n$ being integer and $T$ the temperature. The impurity
Green's function maps onto the site-diagonal Green's function of
the original lattice Hamiltonian by the relationship:
\begin{equation}
G_{d\sigma}(i\omega_{n})\equiv G_{ii\sigma}(i\omega_{n}) =
\int_{-\infty}^{\infty}
\frac{\rho_{\sigma}^{(B)}(\varepsilon)d\varepsilon}{i\omega_{n}
-(\varepsilon-\mu)-\Sigma_{\sigma}(i\omega_{n})} \;,
\label{EQ:SELFCONSISTENCY_1}
\end{equation}
where $G_{d\sigma}(i\omega_{n})$ is the Fourier transform of the
impurity Green's function $G_{d\sigma}(\tau-\tau^{\prime})=-\langle
T_{\tau}[d_{\sigma}(\tau)d_{\sigma}^{\dagger}(\tau^{\prime})]\rangle$
determined from the impurity Hamiltonian Eq.~(\ref{EQ:HAMIL_SIAM})
while $G_{ii\sigma}(i\omega_{n})$ is the site-diagonal Green's function
corresponding to the lattice Hamiltonian in Eq.~(\ref{EQ:HAMIL_LATT}),
$\rho_{\sigma}^{(B)}(\varepsilon)$ is the Bloch density of states
(the
$U=0$ density of states of Eq. (1)) and
$\mu$ is the chemical potential. This procedure indicates that
$\Delta_{\sigma}$ in the present context, in contrast to the original SIAM
model~\cite{PWAnderson61},  is not known {\em a priori}. For infinite
dimensional lattices or the Bethe lattice with infinite coordination
number as we are considering here, the self-energy is site-diagonal
and the DMFT becomes exact. In addition, for $z\rightarrow \infty$,
the non-interacting Bloch density of states is semielliptic, which
reduces the self-consistency condition to:
\begin{equation}
\Delta_{\sigma}(i\omega_{n})=t^{2}G_{d\sigma}(i\omega_{n})\;,
\label{EQ:SELFCONSISTENCY_2}
\end{equation}
with $\epsilon_{d\sigma}=-\mu$ for the paramagnetic case. Keeping
the hybridization function spin dependent allows the approach to
also be applied to the ferromagnetic ordering case.

If one is able to solve the SIAM for arbitrary parameters (i.e.,
$\epsilon_{k\sigma}$ and $V_{k\sigma}$ via $\Delta_{\sigma}$), the
lattice model can be solved through self-consistent iteration, by
using the following procedure: First, given a
hybridization function, one solves the SIAM for the impurity Green's
function (or self-energy); Second, through the self-consistency
condition, Eq.~(\ref{EQ:SELFCONSISTENCY_2}), a new hybridization
function is obtained. The procedure is then iterated until the
self-consistency is achieved.

The EOM approach corresponds to a resummation of low-order hopping
process, but requires a decoupling scheme. We have used the one
introduced by Appelbaum and Penn~\cite{JAAppelbaum69} and
Lacroix~\cite{CLacroix81} (APL), which is known to capture the
right qualitative feature of physics at low temperatures. Briefly,
the EOM method consists of differentiating the Green's function
$G_{d\sigma}(\tau-\tau^{\prime})$ with respect to the imaginary
time $\tau$, thereby generating higher-order Green's functions
which must be approximated in order to close the equations for
$G_{d\sigma}(\tau-\tau^{\prime})$. Since the time derivative of a
Heisenberg operator is determined by its commutator with the
Hamiltonian, it follows: \begin{eqnarray} \biggl{(}
-\frac{\partial }{\partial \tau} -\epsilon_{d\sigma} \biggr{)}
G_{d\sigma}(\tau-\tau^{\prime}) &=& \delta(\tau-\tau^{\prime}) + U
G_{d\bar{\sigma}\sigma}(\tau-\tau^{\prime})\nonumber \\ &&+
\sum_{k} V_{k\sigma}^{*} G_{k\sigma}(\tau-\tau^{\prime}) \;.
\label{EQ:EOM_1}
\end{eqnarray}
On the right-hand side of Eq.~(\ref{EQ:EOM_1}), there are two new
Green's functions: $G_{k\sigma}(\tau-\tau^{\prime})= -\langle
T_{\tau}
[c_{k\sigma}(\tau)d_{\sigma}^{\dagger}(\tau^{\prime})]\rangle $,
and the two-particle Green's function $G_{d\bar{\sigma}\sigma}=-\langle
T_{\tau}[n_{d\bar{\sigma}}(\tau)
d_{\sigma}(\tau)d_{\sigma}^{\dagger}(\tau^{\prime})]\rangle $.
By
using the APL decoupling scheme to truncate this EOM one obtains:
\widetext
\begin{equation}
G_{d\sigma}(i\omega_{n}) = \frac{1-\langle n_{d\bar{\sigma}} \rangle
} {i\omega_{n}-\epsilon_{d\sigma} -\Delta_{\sigma} +\frac{U
\Sigma_{1\sigma}} {i\omega_{n}-\epsilon_{d\sigma}-U -\Delta_{\sigma}
-\Sigma_{3\sigma}}} +\frac{\langle
n_{d\bar{\sigma}}\rangle}{i\omega_{n}-\epsilon_{d\sigma}-\Delta_{\sigma}
-U
-\frac{U\Sigma_{2\sigma}}{i\omega_{n}-\epsilon_{d\sigma}-\Delta_{\sigma}
-\Sigma_{3\sigma}} }\;,
\label{EQ:GREEN}
\end{equation}
\endwidetext
\noindent where the occupancy probability is
\begin{equation}
\langle n_{d\sigma} \rangle =\int d\epsilon f_{FD}(\epsilon)
\rho_{\sigma}(\epsilon) \;,
\label{EQ:OCCUPANCY}
\end{equation}
with the spectral density $\rho_{\sigma}(\epsilon)=
-\mbox{Im}[G_{d\sigma}(\epsilon + i\eta)]/\pi$ ($\eta$ is an
infinitesimal). The hybridization function $\Delta_{\sigma}$
represents the hopping of the electron with spin projection
$\sigma$ onto or out of the impurity site, while the other
self-energies are due to the hopping of the electron with opposite
spin projection $\bar{\sigma}$, \widetext
\begin{equation}
\Sigma_{i\sigma}(i\omega_{n}) = \sum_{k} A_{k\bar{\sigma}}^{(i)}
\vert V_{k\bar{\sigma}} \vert^{2}
\biggl{[} \frac{1}{i\omega_{n} -\epsilon_{d\sigma} +
\epsilon_{d\bar{\sigma}} -\epsilon_{k\bar{\sigma}}}
+\frac{1}{i\omega_{n} -\epsilon_{d\sigma} -\epsilon_{d\bar{\sigma}}
-U + \epsilon_{k\bar{\sigma}}}\biggr{]},\; i=1,2,3\;,
\label{EQ:SELFENERGY}
\end{equation}
\endwidetext
\noindent where
$A_{k\bar{\sigma}}^{(1)}=f_{FD}(\epsilon_{k\bar{\sigma}})$,
$A_{k\bar{\sigma}}^{(2)}=1-f_{FD}(\epsilon_{k\bar{\sigma}})$,
$A_{k\bar{\sigma}}^{(3)}=1$ with the Fermi distribution function
$f_{FD}(\epsilon)=[1+\exp(\epsilon/T)]^{-1}$. By performing an
analytic continuation, i.e., replacing $i\omega_{n}$ with
$\omega + i\eta$, one arrives at the retarded Green's function,
which is exactly identical to that derived directly from the real
time approach~\cite{JAAppelbaum69,CLacroix81,YMeir91}. The impurity Green's
function $G_{d\sigma}$, as given by Eq.~(\ref{EQ:GREEN}), is exact
both in the non-interacting limit ($U=0$) and in the atomic limit
($V_{k\sigma}=0$).  Since the accuracy of the EOM method depends on
how
good the decoupling scheme is, this is an important confirmation
on
the quality of the APL decoupling.
It is also useful to note that at low temperatures, the EOM
solution based on the APL decoupling scheme has the expected Kondo resonance
at the Fermi energy~\cite{TKNg88}, originating from the
self-energy, $\Sigma_{1\sigma}$, which is due to the virtual intermediate
process where the impurity orbital is occupied by an electron of
opposite spin $\bar{\sigma}$.

Physical quantities like the spectral function require a knowledge of
the Green's function on the real frequency axis. For IPT,
direct
calculations of the self-energy
on the real frequency axis are very
complicated because of
multiple-dimension integrals.
Instead one first performs the calculation on the imaginary frequency
axis and then computes the spectral function by performing an analytic
continuation by a Pade approximation. For this kind of
treatment to be successful the quantities involved in the
Green's functions and the self-energies need to be expressed in terms of
the hybridization function as a whole. In the current approach, the
three self-energies in Eq.~(\ref{EQ:SELFENERGY}) cannot be
represented as functions of $\Delta_{\sigma}$. However, by
introducing a
line-width function \begin{equation}
\Gamma_{\sigma}(\omega)=-\frac{1}{\pi}\mbox{Im}[\Delta_{\sigma}(\omega+
i\eta)]\;, \label{EQ:GAMMA}
\end{equation}
one can instead rewrite Eq.~(\ref{EQ:SELFENERGY}) as: \widetext
\begin{equation}
\Sigma_{i\sigma}(i\omega_{n}) = \int d\epsilon A^{(i)}(\epsilon)
\Gamma_{\bar{\sigma}}(\epsilon) \biggl{[} \frac{1}{i\omega_{n}
-\epsilon_{d\sigma} + \epsilon_{d\bar{\sigma}} -\epsilon}
-\frac{1}{-i\omega_{n} +\epsilon_{d\sigma} +\epsilon_{d\bar{\sigma}}
+U - \epsilon}\biggr{]},\; i=1,2,3\;. \label{EQ:SELFENERGY_NEW}
\end{equation}
\endwidetext

\begin{figure}[th]
\centerline{\psfig{file=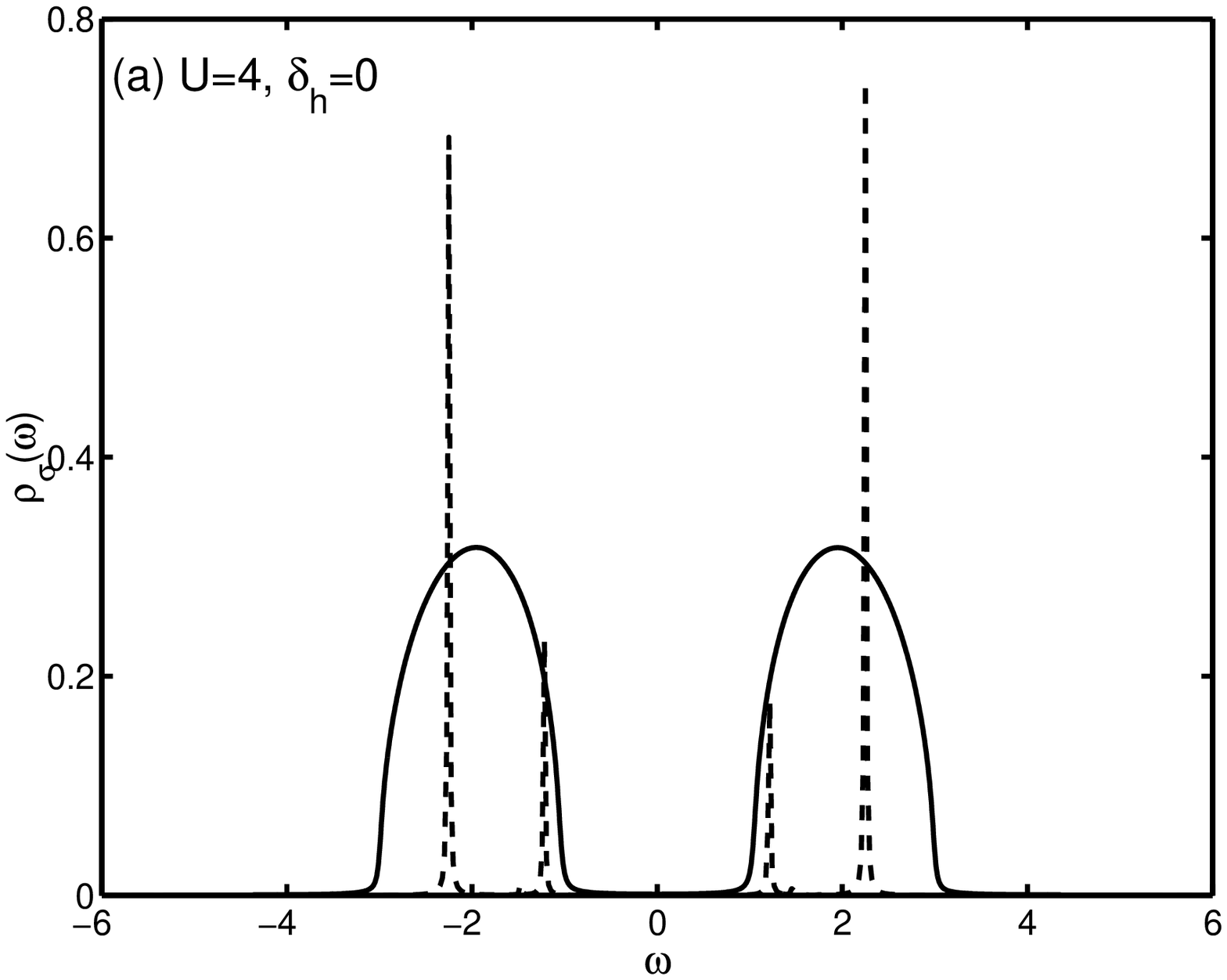,width=6cm}}
\centerline{\psfig{file=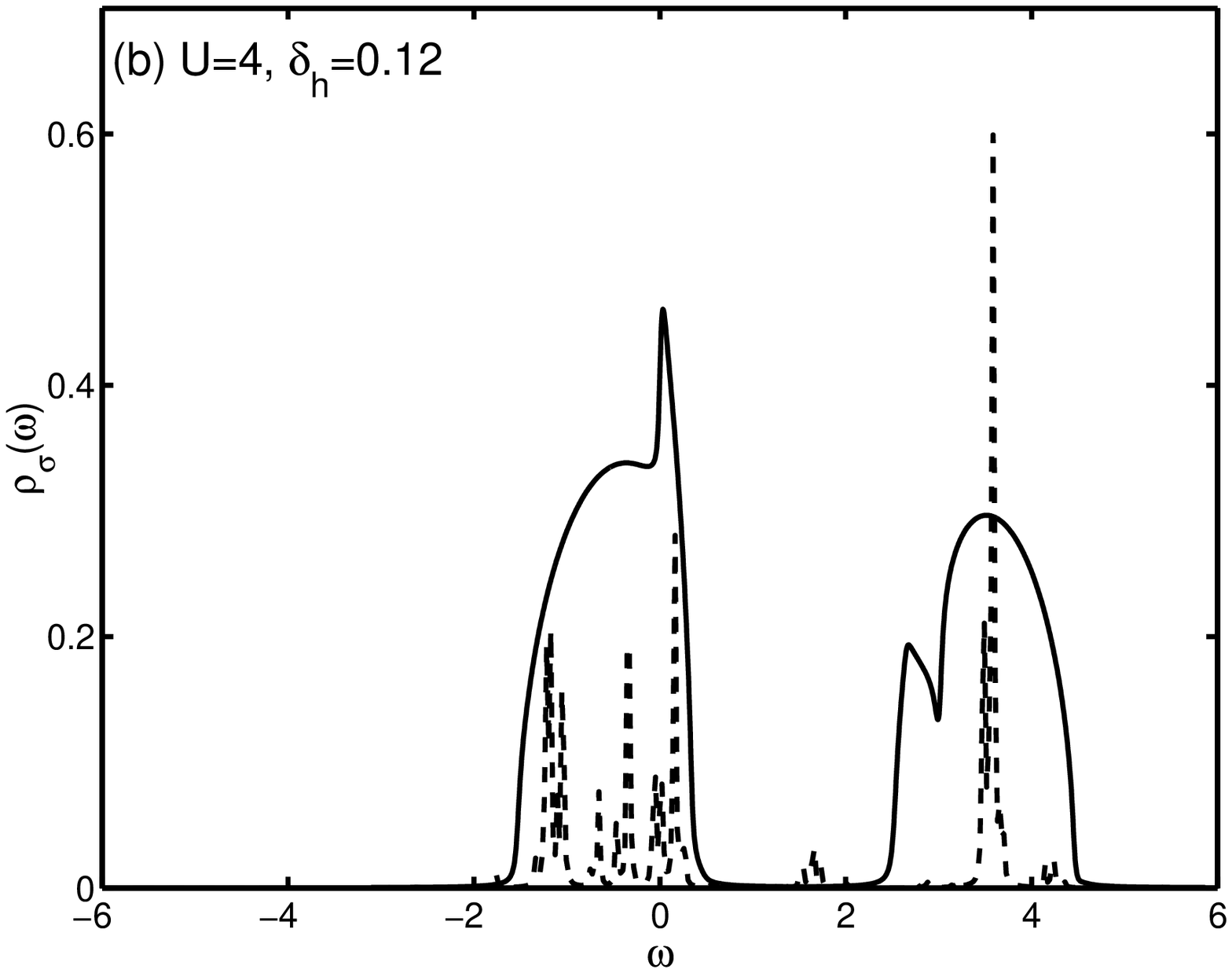,width=6cm}} \caption{The
spectral density as a function of energy for $U=4$ and the hole
doping $\delta_h=0$ (a) and 0.12 (b). Thick solid line: EOM
approach. Thin dashed line: ED (6 sites). For clarity, the
intensity from the ED is scaled down by a factor of 16 (a) and 8
(b). The parameter $\eta=0.01$.  $\omega=0$ corresponds to the
Fermi
energy.} \label{FIG:DOS_1}
\end{figure}

In order to work on the imaginary frequency
axis, at each iteration step a Pade approximation for
$\Gamma_{\sigma}$ would be required,
which would introduce
significant numerical errors and
make the convergence difficult. Since the occupancy probability given by
Eq.~(\ref{EQ:OCCUPANCY}) and the self-energies given by
Eq.~(\ref{EQ:SELFENERGY_NEW}) all involve only one-dimensional
integrals, we find it is more efficient to implement
the technique directly on the real-frequency axis. For this purpose,
we divide the relevant energy range into segments and use the
Simpson's rule in combination with the Neville's polynomial
interpolation algorithm to do these integrals~\cite{WHPress92}. In
the actual calculation, we use two iteration loops: The inner loop is
for the occupancy propability $\langle n_{d\sigma} \rangle$ for a
fixed line-width function $\Gamma_{\sigma}$. After the convergence of
the inner loop is achieved, $\Gamma_{\sigma}$ is updated through
Eqs.~(\ref{EQ:SELFCONSISTENCY_2}) and (\ref{EQ:GAMMA}) in the outer
loop. We find the algorithm is very efficient: in most cases a
solution is found
within 20 iterations. It is also remarkable that the
algorithm provides stable convergence with  respect to different
guesses for the
initial hybridization function used to start the
calculation.

In the following discussions, all energies are measured in units
of the half bandwidth $D=2t=1$. We take the temperature $T=0.01$.
In Fig.~\ref{FIG:DOS_1}, we show the spectral density,
$\rho_{\sigma}(\omega)$, as a function of energy for $U=4$ and for
values of the
hole doping $\delta_h$ ($=1-\sum_{\sigma} \langle n_{d\sigma}
\rangle$) of 0 and 0.12. As shown in
Fig.~\ref{FIG:DOS_1}(a), in the undoped case (i.e., half filling),
the spectral density displays only the lower and upper Hubbard
bands separated by a Mott insulator gap. Due to the particle-hole
symmetry, these two bands are symmetric to each other around the
Fermi energy. At finite doping (e.g., $\delta=0.12$ as shown in
Fig.~\ref{FIG:DOS_1}(b)), the spectral density exhibits a sharp
quasiparticle resonance peak at the Fermi energy, and two broad
satellite features corresponding to the lower and upper Hubbard
bands, as would be expected for a doped Mott insulator. Since
the particle-hole symmetry is broken upon doping, the spectral
density is no longer symmetric with respect to the Fermi energy.
To demonstrate the accuracy of the present approach, we have also
run the ED algorithm developed by Caffarel and
Krauth~\cite{MCaffarel94} for 6 sites with the same parameter
values. The output is shown by thin dashed lines in
Fig.~\ref{FIG:DOS_1}. As one can see, the results obtained from
the EOM approach compare well with those from the exact
diagonalization method. Like IPT, in
addition to computational efficiency, the advantage of
the EOM approach is that it can give a very smooth shape for
the spectral density. The ED method can at best give a correct
overall spectral density distribution with sharp spike-like structures due to
the small number of orbitals being used. Because the use of a
large number of orbtals in the ED method would exceed current
computational capabilities, we conclude that the EOM approach may
be a useful alternative.

\begin{figure}[th]
\centerline{\psfig{file=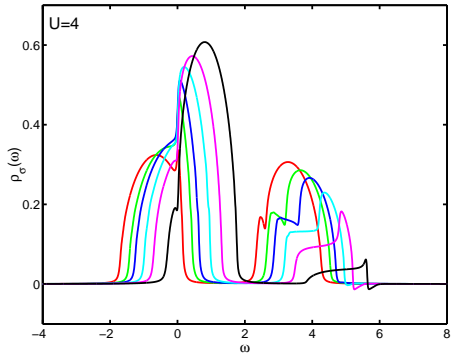,width=6cm}} \caption{(Color) The
evolution of the spectral density with increasing hole doping
$\delta_h=0.06$, 0.16, 0.26, 0.44, 0.64, 0.88 (counted from the
left). The parameter $\eta=0.02$. } \label{FIG:DOS_2} \end{figure}

With increased confidence due to our success in calculating the
spectral function, we now turn to the evolution of the spectral
density with hole doping. These results are shown in
Fig.~\ref{FIG:DOS_2}. The most striking features are as follows:
When the chemical potential crosses the edge of the lower Hubbard
band (i.e., when hole doping begins), a Kondo resonance peak
shows up at the Fermi energy. At the same time, an anti-resonance
dip occurs in the upper band. As the doping increasing, there
continues
to be a peak-like feature near the Fermi energy that
gradually decreases
in intensity.  The effects on the upper Hubbard
band are even more
dramatic, as the
anti-resonance dip evolves into a
feature that shifts more deeply into the
upper Hubbard band and eventually becomes more of
a plateau, with the spectral weight continuing to be transferred
out of the upper Hubbard band. The hole doping dependence of the
spectral weight reflects the different parameter regimes of the
SIAM. Within the EOM approach, one does not find a narrowing of
the upper Hubbard band. This picture of the spectral density
evolution is different from that based on the iterative
perturbation theory. Since the two approaches are so different, we
find it very difficult to draw a judgement on which picture is
more reasonable without a large-size ED calculation to check the
results. We have also studied the electron doping case, and found
a similar spectral weight transfer (now out of the lower Hubbard
band).

To summarize, we have proposed an EOM approach as an impurity
solver for dynamical mean field theory. The solution for the
impurity Green's function is exact in the non-interacting and
isolated-site limits. The technique has been implemented directly
on the real-frequency axis, which turns out to be computationally
efficient. As an illustration, we have studied a finite-$U$
Hubbard model defined on the Bethe lattice with infinite
connectivity at arbitrary filling. Depending on the filling, all
typical features of strongly correlated materials such as the
quasiparticle peak at the Fermi energy, and lower and upper
Hubbard bands, have been produced for the spectral density. The
main results have also been compared with exact diagonalization
and found in good agreement. In addition, a new picture of the
spectral density evolution as a function of filling factor is
found, which should be tested by future exact diagonalization
studies with a large number of orbitals. The extension of this
approach to the multiorbital case constitutes a future
investigation, which will be very useful for {\em ab initio} DMFT
applications to real materials.

{\bf Note Added:} In the final stage of this work, we became aware of
a preprint by H. O. Jeschke and G. Kotliar~\cite{Jeschke04}, where a
similar approach was proposed. Their study focused on the case of an
infinite Hubbard
interaction such that the technique could be directly implemented in
Matsubra frequency. In that case, only the quasiparticle peak and
lower Hubbard band are the relevant characteristics. It is
encouraging to see that
in the comparable region, both of our studies show a very similar
shape for the spectral density.

{\bf Acknowledgments:} One of us (JXZ) is grateful to Q. Si for
helpful discussions. This research is supported by the Department
of Energy under contract W-7405-ENG-36.

\end{document}